\documentclass[pre,twocolumn]{revtex4-1} 

\usepackage{amsmath}  
\usepackage{amsfonts} 
\usepackage{graphicx} 

\graphicspath{{figures/}}  

\let\Gamma\varGamma
\let\Theta\varTheta
\let\Lambda\varLambda
\let\Xi\varXi
\let\Pi\varPi
\let\Sigma\varSigma
\let\Upsilon\varUpsilon
\let\Phi\varPhi
\let\Psi\varPsi
\let\Omega\varOmega

\usepackage{color} 
\definecolor{nc}{rgb}{0.1,0.1,0.5} 

\begin{document}

\title{Kinetic models of immediate exchange}

\author{Els Heinsalu$^{\rm (a,b)}$ and Marco Patriarca$^{\rm (b)}$\\
{\small (a) Niels Bohr International Academy, Niels Bohr Institute, Blegdamsvej 17, DK-2100 Copenhagen, Denmark}\\
{\small (b) NICPB--National Institute of Chemical Physics and Biophysics, R\"avala 10, Tallinn 15042, Estonia}\\
{\it E-mails}: {\tt els.heinsalu@kbfi.ee, marco.patriarca@kbfi.ee}}




\begin{abstract}
We propose a novel kinetic exchange model differing from previous ones in two main aspects.
First, the basic dynamics is modified in order to represent economies where immediate wealth exchanges are carried out, 
instead of reshufflings or uni-directional movements of wealth.
Such dynamics produces wealth distributions that describe more faithfully real data at small values of wealth.
Secondly, a general probabilistic trading criterion is introduced, 
so that two economic units can decide independently whether to trade or not depending on their profit.
It is found that the type of the equilibrium wealth distribution is the same
for a large class of trading criteria formulated in a symmetrical way with respect to the two interacting units.
This establishes unexpected links between and provides a microscopic foundations 
of various kinetic exchange models in which the existence of a saving propensity is postulated.
We also study the generalized heterogeneous version of the model in which units use different trading criteria and
show that suitable sets of diversified parameter values  with a moderate level of heterogeneity can reproduce realistic wealth distributions with a Pareto power law.
\end{abstract}

\maketitle

\newpage

\section{Introduction}

Kinetic exchange models provide a minimal description of wealth exchange between economic units 
(representing, e.g., individuals, families, or companies)
in a way formally similar to that in which energy is transferred between molecules of a fluid due to collisions~\cite{Patriarca2010b,Patriarca2013a}.
Such models were introduced independently in different fields such as social
sciences~\cite{Angle1986a,Angle1992a,Angle2006a}, economics~\cite{Bennati1988a,Bennati1988b,Bennati1993a}, and physics~\cite{Dragulescu2000a,Dragulescu2001a,Dragulescu2001b,Yakovenko2009a,Chakraborti2000a,Chakraborti2002a}.
John Angle~\cite{Angle1986a,Angle1992a} originally introduced this
type of models basing them on the surplus theory, with the goal of
describing the origin of wealth inequalities.
Compared with other agent-based models of financial markets~\cite{Samanidou2007a}, the structure of kinetic exchange models is extremely simple: 
they only describe wealth flows between economic units without considering (explicitly) other elements of a market dynamics.
However, even so they predict quite realistic shapes of wealth distributions~\cite{Chatterjee2005b}.

In kinetic exchange models it is assumed that the total wealth of the
system is conserved, following in turn from wealth conservation 
during each unit-unit interaction.
The dynamics can be formulated through the following update
rule at a generic time iteration $t$ defining the wealth exchange
between two units $j$ and $k$  chosen randomly among the
$N$ units of the system ($j,k = 1, \dots, N$),
\begin{equation} 
\begin{aligned} \label{general-rule}
x_j^{'} = x_j + \Delta x_{jk} \, , \\
x_k^{'} = x_k - \Delta x_{jk} \, .
\end{aligned}
\end{equation}
Here $x_j$ and $x_k$ ($x_j^{'}$ and $x_k^{'}$) are the wealths of the trading
units $j$ and $k$ before (after) the exchange and 
$\Delta x_{jk} = - \Delta x_{kj}$ is the exchanged amount of wealth.
Without loss of generality, in Eqs.~(\ref{general-rule}) the plus (minus) sign 
has been chosen for $j$ ($k$); who wins or loses
depends on the sign of $\Delta x_{jk}$.
The conservation of the total wealth of the system leads to an
equilibrium wealth distribution that coincides formally with the Gibbs
energy distribution.

As discussed in Ref.~\cite{Patriarca2010b}, most of kinetic exchange
models can be formulated in a unified way, where the amount of exchanged wealth in Eqs.~(\ref{general-rule}) is expressed as
\begin{equation} \label{w_exch}
\Delta x_{jk} = \tilde{\omega}_{jk} \, x_k - \tilde{\omega}_{kj} \, x_j  \, .
\end{equation}
Here $\tilde{\omega}_{jk}$ and $\tilde{\omega}_{kj}$ are two suitable
stochastic variables in $(0,1)$ to be extracted at each trade, 
representing the fractions of wealth of unit $k$ and $j$,
respectively, involved in the trade.

Various aspects of kinetic exchange models have been criticized.
A first objection concerns wealth conservation.
Namely, it is pointed out that the assumption of a perfect conservation of wealth
is incompatible with the fact that in real systems wealth is not conserved,
a most relevant reason being the existence of production-consumption processes.
In fact, kinetic exchange models can be interpreted as models with homogeneous production and consumption.
To show this, one can use a modified form of a kinetic exchange model where production and consumption have
been explicitly added to the dynamics.
It is convenient to start from Eqs.~(\ref{general-rule}) and (\ref{w_exch})
and write the total wealth variation of a generic unit $j$
during a time step $\delta t$, defined as the time interval corresponding to one Monte Carlo swap 
over all the units $k$ ($k \ne j$),
\begin{equation} \label{general-rule1}
\frac{x_j' - x_j}{\delta t} \approx \sum_{k (k \ne j)} [ J_{jk} x_k  - J_{kj} x_j ] + (p - c) x_j  \, .
\end{equation}
Here the homogeneous production and consumption terms have been added,
with production and consumption rates $p$ and $c$, respectively; 
$J_{jk} = \tilde{\omega}_{jk}/\delta t$ (and analogously for $J_{kj}$).
In the continuous time limit,
$(x_j' - x_j)/\delta t \approx dx_j(t)/dt$,
one can introduce the auxiliary variables
\begin{equation} \label{XY}
X_j(t) = x_j(t) \exp \! \int \! dt [p(t) - c(t)] \, ,
\end{equation}
where a possible time dependence of  $p$ and $c$ has been taken into account.
Equations~(\ref{general-rule1}) written in the new variables $X_j$ 
do not contain anymore the production and consumptions terms and turn into the equations 
of the corresponding model {\it without} production and consumption,
in which wealth conservation holds.
However, this is not valid in the case of strong heterogeneity, in which
the consumption and production processes of each unit have to be taken into account separately.

Another critic toward kinetic exchange models claims that the type of
trade dynamics they model can hardly resemble an actual economic trade.
In fact, traditional approaches of economics assume that the decisions to carry out a trade
are taken by rational agents or by economic agents with bounded rationality on the
base of the total or partial information available about the system.
Instead, kinetic exchange models may resemble at first sight more a hazard game,
due to their random dynamics~\cite{Hayes2002a,Lux2005a}.
It is clear that due to their statistical nature kinetic exchange models certainly 
do not provide a direct picture of the economic trading or the exchange activity that they are supposed to describe.
So far a satisfactory microscopic picture and justifications of the models are missing.
Their justification is pragmatic in nature,  
in that it relies a posteriori on the successful prediction of realistic shapes of wealth distributions.
A notable exception is the investigation reported in Ref.~\cite{Chakrabarti2009a},
where a direct link with micro-economics is suggested.

The main goal of the present paper is to make a step toward a microscopic foundation of kinetic wealth exchange models.
This is carried out along a twofold path.
As a first element of the proposed reformulation, 
a novel ``immediate-exchange'' dynamics is introduced in Sect.~\ref{immediate}
in order to represent an actual exchange of goods rather than
a random reshuffling or a uni-directional flow of the wealths of the interacting units,
characteristic of other kinetic exchange models.
This novel dynamics is shown to produce equilibrium wealth distributions that better describes empirical data
also in the very small wealth range.
Secondly, in Sect.~\ref{trading} we introduce a probabilistic criterion
that is used by each economic unit in order to decide  whether to carry out an exchange or not.
The exchange is carried out only if both units accept to do it.
Such an acceptance criterion describes the decision process of a single unit during each interaction and can be suitably customized 
to represent the type and amount of partial information available to a unit during the trade.
In the present work we follow a microscopic approach considering acceptance criteria only based 
on the information concerning the ongoing trade,
that is directly available to the two interacting economic units.
To the best of our knowledge a probabilistic approach based on the (partial) information available 
about the quality and price of a product has been introduced in micro-economic models in Refs.~\cite{Lu2008a,Liao2014a},
but the probabilistic criteria used in many-agents models
usually follow an approach based on a utility function~\cite{Miceli2013a}.

As an important check versus real data, the heterogeneous version of the new model is studied in
Sect.~\ref{heterogeneous}.
It is shown that, analogously to e.g. the Chakrabarti-Chakraborti kinetic exchange model,
there exist suitable sets of diversified parameters reproducing realistic wealth distributions on all scales 
including the large wealth range where the Pareto power-law is observed.

Results and possible future lines of research are discussed in the Conclusion.

\section{Immediate-exchange model}
\label{immediate}

\subsection{Formulation of the immediate-exchange model}

In this section a kinetic exchange model that provides a description of a market where encounters between units are accompanied by
immediate exchanges, is proposed.
By ``immediate exchange'' it is meant for example the type of interchanges that characterize barter where goods are directly exchanged without using a medium of exchange, or market economies where goods are exchanged with
money immediately or according to some agreed time schedule. 
Immediate exchanges are to be contrasted with the ``delayed exchanges'' characterizing e.g. gift economies~\cite{Mauss1950a}, 
where for cultural reasons valuables are given without an explicit agreement for immediate or future rewards.
In the latter case one can talk about unidirectional trades.
Unidirectional trades can also be used to describe, e.g., insurance business.

For the sake of clarity, we start by considering a barter model, 
in which one can think of each unit $i$ ($i = 1, 2, \ldots, N$) as having some items  that he is willing to change for something
else at some point; the total value of these items is $x_i$.
The model is evolved in time by extracting randomly two units $j$ and
$k$ at every time iteration.
The two units interchange something with value $\epsilon _j x_j$ and
$\epsilon _k x_k$, respectively, where $\epsilon _{j}$ and $\epsilon_{k}$ are two independent uniform random numbers in the
interval $(0, 1]$, different at each iteration.
The random nature of these quantities describes the situation in which
each time when somebody is trying to do a trade it can be of a different object among all the things he owns and is willing to exchange for something else.

Notice that the model does not distinguish between different types of goods 
but focuses only on the corresponding amounts of wealth exchanged between
units; therefore, one of the goods can be possibly understood as the currency in use.
Thus, the model can be reinterpreted as the one of a currency-based market.
In the case of a currency-based trade, where an object of value $x_j$ is exchanged with some money $x_k$, 
the random numbers describe intrinsic fluctuations of prices and currency values.

The dynamics of the proposed model can be defined by the following equations:
\begin{equation}
\begin{aligned} \label{xj}
x_j^{'} &= (1 - \epsilon_j) x_j + \epsilon_k x_k \, , \\
x_k^{'} &= (1 - \epsilon_k) x_k + \epsilon_j x_j  \, ,
\end{aligned}
\end{equation}
which can be rewritten in the same form of Eqs.~(\ref{general-rule}) with
\begin{equation} \label{Delta}
\Delta x_{jk} = \epsilon_k x_k - \epsilon_j x_j \, . 
\end{equation}
Because $\epsilon_j, \epsilon_k > 0$ then from Eqs.~(\ref{general-rule}) together with (\ref{Delta}) it is clear that
a situation where $x_i = X$ (i.e., a unit $i$ owns all the wealth of the system) or $x_i = 0$ (i.e., a unit $i$ is totally poor) 
is not possible at any moment of time.
In fact, numerical simulations show that the equilibrium distribution of wealth is a $\Gamma$-distribution,
\begin{equation} \label{Gamma0}
f_{\alpha,\beta}(x) = \frac{\beta}{\Gamma(\alpha)} (\beta x)^{\alpha -1} \exp(- \beta x ) \, ,
\end{equation}
with $\alpha = \beta = 2$.
In the following, using the fact that $\langle x \rangle = \alpha/\beta$ 
and the average wealth is constant and always set here as $\langle x \rangle = 1$ due to the initial conditions $x_i = 1$,
one has $\alpha = \beta$ and it is convenient to use the simplified one-parameter form 
$f_\alpha(x) \equiv [f_{\alpha,\alpha/\langle x \rangle}(x)]_{\langle x \rangle = 1}$, i.e.,
\begin{equation} \label{Gamma}
f_\alpha(x) = \frac{\alpha^\alpha x^{\alpha -1}}{\Gamma(\alpha)} \exp(- \alpha x ) \, .
\end{equation}
The equilibrium wealth distribution corresponding to the value of the shape parameter $\alpha = 2$
found from the fitting for the immediate-exchange model is $f_2(x) = 4x\exp(-2x)$.
Thus, the wealth distribution is zero at zero wealth, $f_2(0) = 0$, and decays exponentially at large $x$. 

It is to be noticed that the value of $\alpha \approx 2$ is in the range $(2,2.5)$ of the values characterizing real distributions, 
such as  the household incomes analyzed by Salem and Mount~\cite{Salem1974a,Angle1986a}.
This value has been obtained using the plain immediate-exchange dynamics without further hypotheses,
as for instance the introduction of an explicit saving propensity constraining the amount of wealth entering an exchange,
as in the Chakraborti-Chakrabarti model.

As a technical remark, the considerations above are valid for
random numbers $\epsilon_i$ uniformly distributed in $(0,1)$. 
Employing random numbers with different distributions will produce accordingly a
modified shape of the equilibrium wealth distribution.
Details about the numerical algorithms used in the simulations of kinetic exchange models 
can be found in Ref.~\cite{Patriarca2013a}.

We also notice that the dynamics described by Eqs.~(\ref{xj})
is in the very spirit of kinetic exchange models, reflecting even more closely than other kinetic exchange models the statistical similarity of wealth flows to inter-molecular energy exchanges.
In fact, according to kinetic theory, the energy exchanged in a collision between two molecules $j$ and $k$ has the form given by Eq.~(\ref{Delta}). 
The variables  $\epsilon_j$ and $\epsilon_k$ depend then on the initial directions of the
molecular velocities and are to be considered \emph{independent random
numbers} in the hypothesis of molecular chaos.
For further details see Ref.~\cite{Chakraborti2008a}.

\subsection{Immediate-exchange versus unidirectional and reshuffling models}
\label{compare}

The immediate-exchange model introduced above is formally very similar to a model
introduced by Dr$\breve{\mathrm{a}}$gulescu and Yakovenko \cite{Dragulescu2000a},
described by Eqs.~(\ref{general-rule}) with
\begin{equation} \label{Delta-DY}
\Delta x_{jk} = \epsilon x_k - (1 - \epsilon) x_j  \, ,
\end{equation}
where $\epsilon$ is a random number in $(0, 1]$.
By comparing Eqs.~(\ref{Delta}) and (\ref{Delta-DY}), one can see that while in the dynamics of the
model of Dr$\breve{\mathrm{a}}$gulescu and Yakovenko only one random number $\epsilon$ is present, the model introduced above
contains two random numbers and reduces to the first one by setting $\epsilon_j = 1 - \epsilon_k$.
However, in the context of wealth exchanges such a strong correlation
between $\epsilon_j$ and $\epsilon_k$ is difficult to understand.
In fact, the dynamics of the  Dr$\breve{\mathrm{a}}$gulescu and Yakovenko model can be
better understood by rewriting the update relations as 
\begin{equation}
\begin{aligned} \label{DY}
x_j^{'} &= \epsilon (x_j + x_k)  \, , \\
x_k^{'} &= (1 - \epsilon) (x_j + x_k) \, ,
\end{aligned}
\end{equation}
representing a random reshuffling in a single time of the total
initial amount $x_j + x_k$ between the two interacting units.

The presence of two independent random numbers $\epsilon_j$ and $\epsilon_k$ \emph{versus} the single random number $\epsilon$ may seem a technical detail, but it implies basically different interpretations. 
Importantly, the model of  Dr$\breve{\mathrm{a}}$gulescu and Yakovenko leads to the
exponential equilibrium wealth distribution describing a society where
most of the people are really poor and the distribution mode is
$\bar{x} = 0$. 
Instead, the model proposed leads to the
$\Gamma$-distribution $f_\alpha(x)$ with shape parameter $\alpha = 2$, 
which corresponds to a society where most of the people have the wealth
around the average value $\langle x \rangle = 1$ (assuming that
initially each unit has a wealth $x = 1$), with a mode $\bar{x} = 1/2$
and there is nobody with $x = 0$.

The model of  Dr$\breve{\mathrm{a}}$gulescu and Yakovenko was modified by Chakraborti and
Chakrabarti \cite{Chakraborti2000a} assuming that it is not the total
initial amount $x_j + x_k$ that is reshuffled randomly between the two
interacting units but only a part $(1 - \lambda) (x_j + x_k)$,
while a fraction $\lambda$ is put aside.
The corresponding exchange rule reads:
\begin{equation} 
\begin{aligned} \label{C-model}
x_j^{'} &= \lambda x_j + \epsilon (1 - \lambda) (x_j + x_k)  \, , \\
x_k^{'} &= \lambda x_k + (1 - \epsilon) (1 - \lambda) (x_j + x_k) \, .
\end{aligned}
\end{equation}
The latter equations are equivalent to Eqs.~(\ref{general-rule}) with
\begin{equation} \label{C-model-Delta}
\Delta x_{jk} = (1 - \lambda) [(1 - \epsilon) x_j + \epsilon x_k]  \, .
\end{equation}
In this model the equilibrium distribution of wealth is well described
by a $\Gamma$-distribution with an $\alpha > 1$, given by \cite{Patriarca2004b,Patriarca2004a,Patriarca2005a,Apenko2014a} 
\begin{equation}
\label{alpha0}
\alpha = \frac{1 + 2 \lambda}{1 - \lambda} 
\end{equation}
if $\langle x \rangle = 1$.
This implies that $f(x \!=\! 0) = 0$ and a mode $\bar{x} > 0$, if $\lambda > 0$.

Thus, the exponential shape of the equilibrium wealth distribution of the Dr$\breve{\mathrm{a}}$gulescu and Yakovenko model 
arises from the possibility that during a given interaction a unit can in principle lose \textit{all} wealth that will go to some other unit.
Considering Eqs.~(\ref{DY}), 
this will happen to unit $j$ whenever the extracted value of the random number $\epsilon$ is close enough to zero. 
The model of Chakraborti and Chakrabarti~\cite{Chakraborti2000a} is in
this respect illuminating, since it shows that due to the introduction of saving (through a saving propensity $\lambda > 0$)
such a situation never occurs and there are no units for $x \to 0$, leading not to the exponential 
but to a $\Gamma$-distribution (\ref{Gamma}) with shape parameter (\ref{alpha0}).
Therefore, units carrying out immediate exchanges can be formally seen as equivalent to units with a saving propensity.
More precisely the immediate exchange model proposed leads to the same equilibrium wealth distribution 
as the model of Chakraborti and Chakrabarti with $\lambda = 1/4$.
The fundamental difference is that in the model proposed here, the situation with $x_i = 0$ is excluded naturally without further
assumptions solely by the fact that even if during an exchange a unit $j$ gives away everything, 
he always receives something in exchange from the other unit $k$ and thus, after the transaction, 
one always finds an $x_i > 0$ for each $i$.


\begin{figure}[ht]
\resizebox{0.45\textwidth}{!}{
  \includegraphics{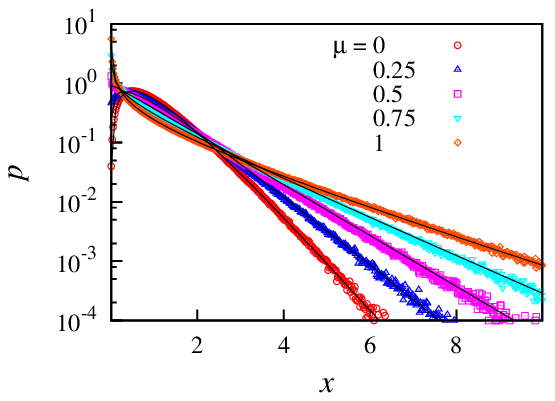}
}
\resizebox{0.45\textwidth}{!}{
  \includegraphics{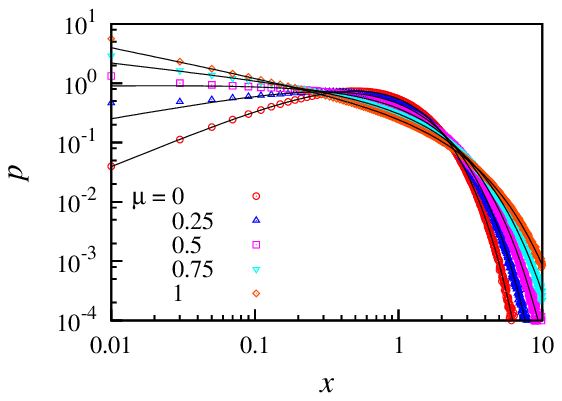}
}
\caption{\label{Fig_mu}
Semilogarithmic (top) and logarithmic (bottom) plot of the equilibrium distribution
$p(x)$ of wealth $x$ for different values of $\mu$, i.e., for
different fractions of unidirectional interactions.
The value $\mu = 0$ corresponds to immediate exchanges only and
$\mu = 1$ to the situation when all the interactions are unidirectional.
Dots represent results from numerical simulations, while lines are a
fitting using the $\Gamma$-distribution (\ref{Gamma}) with $\alpha$
given by Eq.~(\ref{alpha-mu}) for different values of $\mu$.
}
\end{figure}

While the model of Dr$\breve{\mathrm{a}}$gulescu and Yakovenko assumes that the total amount of wealth, $x_j + x_k$, is reshuffled randomly between
the two interacting units, the first model introduced by Angle \cite{Angle1983a,Angle1986a} assumes a unidirectional flow:
with probability $p_0$ a random fraction $\epsilon$ of the wealth $x_j$ of unit $j$ is transferred to unit $k$, while with probability
$1 - p_0$ a random fraction $\epsilon$ of the wealth $x_k$ of unit $k$
is transferred to unit $j$.
Considering the particular case of the symmetrical interaction, i.e., $p_0 = 1/2$, then the dynamical evolution of the model is determined by
Eqs.~(\ref{general-rule}) with
\begin{equation} \label{A-model-Delta}
\Delta x_{jk} = \epsilon [\eta x_j - (1 - \eta) x_k]  \, .
\end{equation}
Here the stochastic variable $\eta$ can assume the values $0$ or $1$
with probability $1/2$; $\epsilon$ maintains the same meaning as above.
In the case of the unidirectional wealth flow model of Angle, the growth of the
fraction of poor units with wealth close to zero is even more dramatic
than in the reshuffling dynamics considered above: in fact, the
resulting equilibrium wealth distribution of this model is found to
be a $\Gamma$-distribution with shape parameter $\alpha = 1/2$~\cite{Katriel2014a},
that \emph{diverges} for $x \to 0$, signaling a wealth accumulation in the
hands of a very few units.

When a saving propensity is introduced in this model, 
assuming that no wealth fraction greater than $(1 - \lambda)$ can be transferred from one unit to the other, then Eq.~(\ref{A-model-Delta}) for the exchanged wealth becomes
\begin{equation} \label{Asave-model-Delta}
\Delta x_{jk} = \epsilon (1 - \lambda) [\eta x_j - (1 - \eta) x_k]  \, .
\end{equation}
In this case, the equilibrium wealth distribution for a generic value
of $\lambda$ is still described by a $\Gamma$-distribution with shape 
parameter~\cite{Patriarca2010b}
\begin{equation}
\label{alpha1}
\alpha = \frac{1 + 2 \lambda}{2(1 - \lambda)} \, ,
\end{equation}
which is just half of the value given by Eq.~(\ref{alpha0}) for the Chakraborti and
Chakrabarti model.
In this case, the divergence at $x \to 0$ will persist as long as
$\alpha < 1$, meaning saving propensities $\lambda < 1/4$.
At $\lambda = 1/4$ the exponential distribution is obtained, while
for $\lambda > 1/4$ the distribution recovers the bell shape with mode
$\bar{x} > 0$.
The value $\alpha = 2$ of the immediate exchange model introduced
above is recovered for $\lambda = 1/2$.

One could expect that (at least in the case of symmetrical interactions) a long series of unidirectional wealth flows 
provides results equivalent to those produced by an analogous series of immediate (bidirectional) exchanges.
However, the basic difference in the region at $x \to 0$ between the shapes of the equilibrium wealth
distributions of the model of Angle and of the immediate-exchange model proposed here
reveal that this is not the case.
A better understanding of the latter issue can be obtained by investigating a system where 
at each time iteration the pair of units extracted for a trade
will carry out with probability $\mu$ the unidirectional transaction
defined by Eq.~(\ref{A-model-Delta}) and with probability $1 - \mu$
the immediate exchange defined by Eq.~(\ref{Delta}).

If $\mu = 0$, i.e., all interactions are immediate exchanges, the
equilibrium wealth distribution is a $\Gamma$-distribution with
$\alpha = 2$.
For any $\mu > 0$ the equilibrium distributions are still well
described for sufficiently large $x$ by a $\Gamma$-distribution, see
Fig.~\ref{Fig_mu}.
The dependence of $\alpha$ on $\mu$ is found to be given by 
\begin{equation} \label{alpha-mu}
\alpha (\mu) = 2 \exp[-\ln(4) \mu] ~ = 2^{1 - 2\mu} \, .
\end{equation}
For $\mu = 1/2$, i.e., when half of the interactions are
unidirectional and half are immediate exchanges, $\alpha = 1$ and the
equilibrium wealth distribution is exponential, $f(x) = \exp (-x)$.
For $x < 0.05$ the distributions deviate from the
$\Gamma$-distributions, see Fig.~\ref{Fig_mu} for $x \to 0$; the
deviation is the larger the larger is the fraction of unidirectional interactions, $\mu$.
For $\mu = 1$ the model described by
Eqs.~(\ref{general-rule}) and (\ref{A-model-Delta}) is recovered.

Here, units making immediate exchanges a fraction $(1 - \mu)$ of times
can be interpreted as units with an effective saving propensity $\tilde{\lambda}(\mu) < \lambda$,
where $\lambda$ is the value of the saving propensity of the Angle model,
corresponding to $\mu = 1$.
The explicit relation between the effective saving propensity $\tilde{\lambda}(\mu)$ 
and the parameter $\mu$ can be found combining Eqs.~(\ref{alpha1}) and (\ref{alpha-mu}),
\begin{equation} \label{mu-lambda-2}
\tilde{\lambda}(\mu) = \frac{ 2^{2(1 - \mu)} - 1 }{ 2^{2(1 - \mu)} + 2 } \, .
\end{equation}
From here one can see that to $\mu = 1$ corresponds $\lambda = 0$,
i.e., the units can lose everything during an exchange. 
Instead, $\mu = 1/2$ leads to the same equilibrium distribution as the
model of Angle with saving propensity $\lambda = 1/4$, and $\mu = 0$
to the equilibrium distribution with saving propensity $\lambda = 1/2$.
Distributions corresponding to values of $\lambda > 1/2 $
cannot be obtained, since they would correspond to negative values of $\mu$.

\section{The influence of a trading criterion}
\label{trading}

\subsection{Formulation of the acceptance criterion}

Let us now introduce a trading criterion on the basis of what each unit will decide whether to make the trade or not
(in other kinetic wealth exchange models the two randomly chosen units always make the trade).
The introduction of probabilistic factors influencing trades is a possible way to go beyond 
the assumption of perfect knowledge of the units assumed in neo-classical economic models.
A probability law suitably describes the natural lack of perfect knowledge of the trading units
concerning the product and its actual value measured as wealth, 
as well as the effect of personal feelings about the goods to be exchanged that can vary from time to time,
or other external random perturbations affecting their decisions.

\begin{figure}[th]
\resizebox{0.45\textwidth}{!}{
  \includegraphics{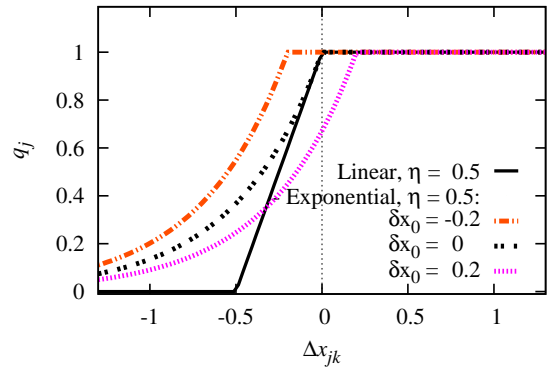}
}
\caption{\label{Fig_p}
Examples of probability $q_j (\Delta x_{jk})$ that unit $j$ accepts to make a
wealth exchange with unit $k$ leading to a net wealth variation $\Delta x_{jk}$ of unit $j$.
Here ``linear'' refers to the piecewise linear probability in
Eq.~(\ref{prob-4tr}) for $\eta = 1/2$, while the other ``exponential''
lines represent the probability given by Eq.~(\ref{prob-5}) for $\eta = 1/2$ and $\delta x_0
= -0.2; 0; 0.2$. 
See text for details.}
\end{figure}

Here, the probabilistic acceptance criterion is assumed to depend only on the information currently available 
to the two trading units at the moment of the trade, i.e.,  
we will use forms of acceptance criterion that depend on the wealths owned by the two interacting units  
before the (possible) trade and on the amount of wealth exchanged.
Without loss of generality, one can focus on the generic unit $j$
that will be assumed to accept to carry out the trade with an acceptance probability $q_j$.
As a simple example of acceptance probability we consider in the following the linear piece-wise function given by
\begin{equation} \label{prob-4tr}
\begin{aligned}
q_j(\Delta x_{jk}) =
\begin{cases}
0 \, , 
&\mathrm{if}~  \Delta x_{jk} < - \eta \langle x \rangle  \, , \\
1 + \Delta x_{jk}/(\eta \langle x \rangle) \, , 
&\mathrm{if}~ - \eta \langle x \rangle < \Delta x_{jk} < 0 \, , \\
1 \, ,          
&\mathrm{if}~ \Delta x_{jk} \ge 0  \, .
\end{cases}
\end{aligned}
\end{equation}
Here $\eta$ is a parameter, $\langle x \rangle = \sum_{i = 0}^N x_i / N$ is the average wealth of the system, and 
$\Delta x_{jk}$ is given by Eq.~(\ref{Delta}). 
This function is depicted as a continuous line in Fig.~\ref{Fig_p}.

In this example, the criterion for accepting or not the exchange depends 
on the quantity $\eta \langle x \rangle$ and on the exchanged wealth $\Delta x_{jk}$.
If $\Delta x_{jk}$ is negative (the unit $j$ is going to lose something) 
with $|\Delta x_{jk}|$ comparable with the scale $\eta \langle x \rangle$, 
then it is highly probable that there will be no trade as
the unit $j$ who would lose is most likely not interested of the exchange. 
The exchange will certainly not take place if $|\Delta x_{jk}| > \eta \langle x \rangle$. 
If $\Delta x_{jk}$ is still negative but with $|\Delta x_{jk}|$ not
too large compared to the scale $\eta \langle x \rangle$ then it is
possible that unit $j$ will accept the exchange anyway, even
though losing something. 
The latter situation could correspond, e.g., to a situation in which 
he is not capable to estimate the loss, or he is simply interested in the
other object for whatever reason and therefore is ready to accept a
limited loss.
For $\Delta x_{jk} > 0$ unit $j$ will always accept the exchange.
Both trading units $j$ and $k$ will independently adopt the same criterion,
the criterion of unit $k$ being defined by an analogous function
$q_k(\Delta x_{kj}) = q_j(-\Delta x_{jk})$ obtained from the function
given by Eq.~(\ref{prob-4tr}) by exchanging $k$ and $j$ and, correspondingly,
$\Delta x_{kj}$ with $-\Delta x_{jk}$.
The decision of each agent $k$ or $j$ is taken probabilistically
with the help of an additional random number that is extracted independently
and compared with $q_j(\Delta x_{jk})$ or $q_k(\Delta x_{kj})$ for the decisions of unit $j$ and $k$, respectively. 
Since the quantity $\Delta x_{jk} = \epsilon_k x_k - \epsilon_j x_j$ 
is given as the difference between the values $\epsilon_k x_k$ and $\epsilon_j x_j$
of the wealth goods exchanged, it represents an actual measure
of the profit or loss of a unit in an exchange.
Thus, the type of interaction between units formulated above
is expected to simulate an actual barter or trade realistically, 
since whether the units will carry out the exchange or not depends on their
corresponding profit or loss.

It is possible to construct similar functions with specific properties and 
depending on additional parameters.
We employ in the following some other forms of the acceptance probability function,
in particular the following exponential shape,
\begin{equation} \label{prob-5}
\begin{aligned}
q_j(\Delta x_{jk}) =
\begin{cases}
\exp[ (\Delta x - \delta x_0) / (\eta \langle x \rangle) ] \, , 
&~\mathrm{if}~ \Delta x_{jk} < \delta x_0 \, , \\
1 \, , 
&~\mathrm{if}~ \Delta x_{jk} \ge \delta x_0 \, .
\end{cases}
\end{aligned}
\end{equation}
This probability function has an exponential shape for $\Delta x_{jk} < \delta x_0$ and is equal to $1$ otherwise.
The shapes of the curve corresponding to an acceptance parameter $\eta = 0.5$ and shift parameters 
$\delta x_0 = 0, \pm 0.2$ are shown in Fig.~\ref{Fig_p}. 
The parameter $\delta x_0$ can be used to make the acceptance criterion stricter or looser 
--- notice that an analogous parameter can in principle be introduced in any other function $q_j(\Delta x_{jk})$.
The acceptance criterion become looser for $\delta x_0 < 0$,
when a unit can accept to carry out an exchange even if accompanied by a limited loss of the order of $|\delta x_0|$.
Instead, the criterion becomes stricter for $\delta x_0 > 0$, since in this case trades are accepted with certainty 
only if the gain is larger than the threshold value $\delta x_0$.

\begin{figure}[ht]
\resizebox{0.45\textwidth}{!}{
  \includegraphics{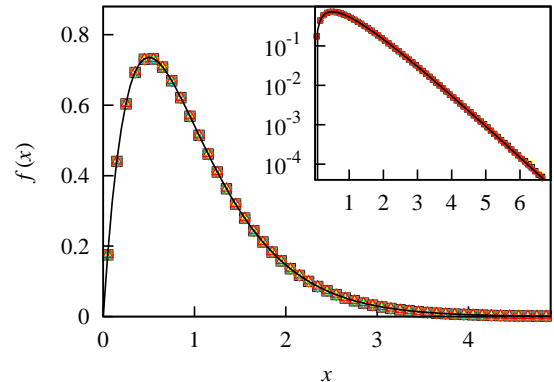}
}
\caption{\label{Fig_eta}
Equilibrium wealth distributions $f(x)$ of the immediate-exchange model.
All the distributions obtained using different or no acceptance criteria (dots) 
collapse on the same $\Gamma$-distribution $f_\alpha(x)$ with $\alpha = 2$ (continuous curve).
The criteria used employ
(a) the ``linear'' probability in Eq.~(\ref{prob-4tr}) with $\eta = 0.1; 0.5; 1; 5; 10;$ and
(b) the ``exponential'' probability in Eq.~(\ref{prob-5}) with $\eta = 0.5$ and $\delta x_0 = -0.4; -0.2; 0; 0.2; 0.4$.
Inset: same plot in semilogarithmic scale.
}
\end{figure}

%
%

\subsection{Acceptance criterion in the immediate exchange model}

We start by applying the acceptance criterion (\ref{prob-4tr}) 
to the kinetic model of immediate exchange introduced in Sect.~\ref{immediate}.
The equilibrium distributions of wealth obtained after adding the
acceptance criterion turns out to be a $\Gamma$-distribution (\ref{Gamma}),
namely the distribution $f_2(x)$ with the same shape parameter $\alpha = 2$
independently of the value of $\eta$.
Thus, the shape of the equilibrium wealth distribution remains unchanged
with respect to the case in which no acceptance criterion is used.

This is true also when different acceptance criteria are employed.
The equilibrium wealth distributions obtained employing
the piece-wise linear function (\ref{prob-4tr}) and 
those corresponding to the exponential acceptance probability function (\ref{prob-5})
for different values of the shift parameter $\delta x_0$
are compared with each other and with the analytical shape of the distribution $f_2(x)$  in Fig.~\ref{Fig_eta}.
 
The value of $\eta$ in Eqs.~(\ref{prob-4tr}) and (\ref{prob-5}) as
well that of $\delta x_0$ in Eq.~(\ref{prob-5}) change the relaxation
process to equilibrium~\cite{Patriarca2007a} (e.g. a smaller $\eta$ corresponds to a larger relaxation time)
but surprisingly do not have any importance for the equilibrium wealth distribution, that is exactly
the same one obtained when assuming that the two randomly chosen units
make the trade with probability $q \equiv 1$.

\subsection{Asymmetrical acceptance criteria}

The considerations above lead to the conclusion that introducing a decision making process 
that can be interpreted as the trial to introduce intelligence in the units behavior, is irrelevant for the final state of the system.
However, this is true only as long as the acceptance criterion is formulated symmetrically with respect to the two trading units.
Different asymmetrical criteria lead to different forms of the equilibrium wealth distribution
and to shapes of the equilibrium wealth distributions different from $\Gamma$-distributions.
Such asymmetrical criteria can be of interest in the study of preferential attachment effects, e.g., when considering
the influence of the richness of the two trading units on the outcome of the trade.
That is the case of the so-called ``rich get richer effect'' already considered in the model of Angle~\cite{Angle1986a},
characterized by uni-directional wealth flows taking place with a higher probability from the poorer to the richer unit.
Notice that unidirectional models represent actual realizations of the so-called ``Matthew effect''~\cite{Merton1968a}, 
since during an encounter not only the richer unit becomes richer with a higher probability but correspondingly, 
due to wealth conservation, the poorer unit becomes even poorer.

As a test, we have studied an asymmetrical version of the immediate exchange
model discussed above, in which a criterion favoring richer units was introduced by always allowing trades with a gain for
the richer trading unit and preventing trades accompanied by a net
gain of the poorer unit a fraction $\theta$ of times, $0 \le \theta \le 1$.

\begin{figure}[ht]
\resizebox{0.45\textwidth}{!}{
  \includegraphics{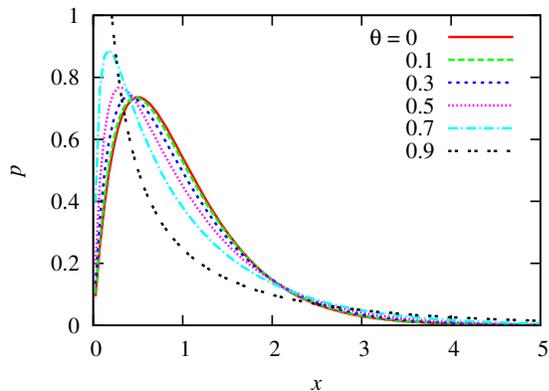}
}
\caption{\label{Fig_theta}
Equilibrium wealth distribution obtained using an asymmetrical
criterion for different fractions $\theta$ of trades that would be
advantageous for the richer unit; see text for details.
Equilibrium distributions with $\theta > 0$ are not well fitted by a
$\Gamma$-distribution (comparison not shown).
}
\end{figure}

In Fig.~\ref{Fig_theta} the equilibrium wealth distribution for some values of $\theta$ ranging between $\theta = 0$ (corresponding to
symmetrical exchanges) and $\theta = 0.9$ (representing a strong
asymmetry in the exchange criterion) are shown.
One can observe how the number of units in the small-$x$ region increases with $\theta$ and that, correspondingly, the distribution mode
shifts leftwards.
It is to be noticed that for high enough values of $\theta$ the distribution diverges for $x \to 0$ --- Fig.~\ref{Fig_theta} shows
such a case for $\theta = 0.9$.
Most importantly, the shapes of the equilibrium distributions obtained for
$\theta > 0$ are not well fitted anymore by a $\Gamma$-distribution
(comparison not shown).

%

\subsection{Acceptance criteria in other models: relative acceptance criteria}

In order to check for other instances of the invariance properties 
found above for the immediate-exchange model,
we have numerically investigated the effects of an acceptance criterion in
other kinetic exchange models, namely in
the Dr$\breve{\mathrm{a}}$gulescu-Yakovenko~\cite{Dragulescu2000a}
and in the Chakraborti-Chakrabarti model with saving propensity~\cite{Chakraborti2000a}.

We start from the Dr$\breve{\mathrm{a}}$gulescu-Yakovenko model, 
defined by Eqs.~(\ref{DY}) or by Eqs.~(\ref{general-rule}) with (\ref{Delta-DY}),
whose equilibrium wealth distribution is a simple exponential shape,
$f(x) = \exp(-x)$ (if $\langle x \rangle = 1$)~\cite{Dragulescu2000a}.
In fact, we find that the exponential form remains unchanged, 
independently of the form of the \emph{symmetrical} acceptance criterion used,
such as those defined by Eqs.~(\ref{prob-4tr}) or (\ref{prob-5}). 
The independence from the acceptance criteria widens the validity range and
justifies the use of the original minimal version of the model
where no acceptance criterion is present,
instead of more elaborate versions where units follow more ``intelligent'' principles.

However, the independence of the equilibrium distribution of a symmetrical acceptance criterion  does not hold for 
the Chakraborti-Chakrabarti model with saving propensity~\cite{Chakraborti2000a}.
In this case one obtains a different equilibrium wealth distribution,
that cannot be fitted anymore by a $\Gamma$-distribution.

\begin{figure}[ht]
\resizebox{0.45\textwidth}{!}{
  \includegraphics{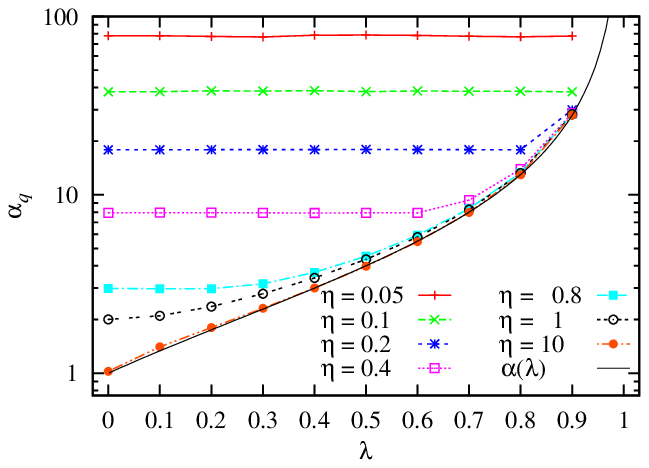}
}
\resizebox{0.45\textwidth}{!}{
  \includegraphics{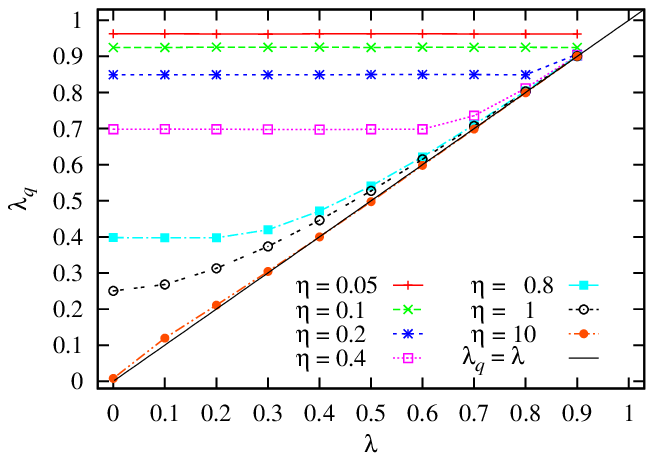}
}
\caption{\label{Fig_R1_a-VS-L}
Fitting of the equilibrium wealth distribution of the Chakraborti-Chakrabarti model 
with acceptance criterion defined by Eq.~(\ref{prob-4-rel}) through a $\Gamma$-distribution.
Top: 
shape parameter $\alpha_q$ {\it versus} saving parameter $\lambda$ for different acceptance parameters $\eta$.
The black solid line is the value of $\alpha(\lambda)$ given by Eq.~(\ref{alpha0}),
corresponding to the Chakraborti-Chakrabarti model,
recovered in the $\eta \to \infty$ limit, i.e., when there is no acceptance criterion.
Bottom: 
effective saving parameter $\lambda_q$ corresponding to the values of $\alpha_q$ in the top panel.
In general $\lambda_q > \lambda$. In the limit of large $\eta$ the presence
of an acceptance criterion becomes negligible and $\lambda_q \to \lambda$.}
\end{figure}

Considering the homogeneity of the update rules of the model, 
we have checked the effect of a probabilistic constraint on the relative variable
 $\Delta x_{jk}/x_j$ of unit $j$ rather than on the absolute wealth exchange $\Delta x_{jk}$,
finding that a good fitting of the equilibrium wealth distributions through a $\Gamma$-distribution is recovered in this case.
In other words, if in the Chakraborti-Chakrabarti model with a given saving propensity $\lambda$
an acceptance criterion is introduced, defined in terms of an (arbitrary) acceptance probability function $Q_j$ 
for each unit $j$  depending on the relative amount of wealth exchanged $\Delta x_{jk}/x_j$,
then the equilibrium wealth distribution is still a $\Gamma$-function, 
but with a shape parameter different from the value given by Eq.~(\ref{alpha0}). 
For clarity we discuss only the acceptance criterion defined by the acceptance probability $Q_j$ obtained by modifying 
the piece-wise linear acceptance probability $q_j$ in Eq.~(\ref{prob-4tr}),
\begin{equation} \label{prob-4-rel}
\begin{aligned}
Q_j(\Delta x_{jk}/x_j) \!=\!
\begin{cases}
0 \, ,                                                 &\mathrm{if}~ \Delta x_{jk}/x_j < - \eta , \\
1 \!+\! \Delta x_{jk}/(\eta \, x_j), &\mathrm{if} ~ -\eta \!<\! \Delta x_{jk}/x_j \!<\! 0, \\
1 \, ,                                                 &\mathrm{if}~ \Delta x_{jk} \ge 0 \, .
\end{cases}
\end{aligned}
\end{equation}
The rescaled variable $\Delta x_{jk}/x_j$ has the new range  $(-1,1)$.
Results are summarized in Fig.~\ref{Fig_R1_a-VS-L}-top, showing the new shape
parameter $\alpha_q$ of the $\Gamma$-distribution $f_{\alpha_q}(x)$ fitting the equilibrium wealth distribution
{\it versus}  the corresponding value of the saving propensity $\lambda$ of the  model,
for different values of the acceptance parameter $\eta$.
In the limit of very large values of $\eta$, in which the acceptance criterion become very loose
and becomes equivalent to no criterion present,
results reduce to those of the original Chakraborti-Chakrabarti model,
represented by the continuous line representing $\alpha(\lambda)$ of Eq.~(\ref{alpha0});
in all other cases one has  $\alpha_q >  \alpha(\lambda)$.

The results can also be expressed by saying that
the relative acceptance criterion with acceptance parameter $\eta$, defined by Eqs.~(\ref{prob-4-rel}), 
turns the equilibrium wealth distribution of the Chakraborti-Chakrabarti model with saving propensity $\lambda$
into that of the {\it same} model with a different saving propensity $\lambda_q$.
The value of the effective saving propensity $\lambda_q$ corresponding to the observed $\alpha_q$ 
can be obtained inverting the relation (\ref{alpha0}) and is given by 
\begin{equation} \label{L_q}
\lambda_q = (\alpha_q - 1)/(\alpha_q + 2) \, .
\end{equation}
For convenience the values of  $\lambda_q$ are depicted in Fig.~\ref{Fig_R1_a-VS-L}-bottom.
It is clear that for every value of the acceptance parameter $\eta$ one has $\lambda_q > \lambda$, 
i.e., the use of an acceptance criterion is equivalent to a larger saving propensity.
For large values of $\eta$, when the acceptance criterion becomes very loose,
results reduce to the original Chakraborti-Chakrabarti model without acceptance criterion
and $\lambda_q \to \lambda$.

A surprising point implied by these results is that the saving propensity
of the Chakraborti-Chakrabarti model can be generated by introducing a relative acceptance criterion.
In fact, the considerations above also apply to the Dr$\breve{\mathrm{a}}$gulescu-Yakovenko model,
that is obtained from the Chakraborti-Chakrabarti model in the limit $\lambda \to 0$.
Thus, one can start from the Dr$\breve{\mathrm{a}}$gulescu-Yakovenko model without saving propensity
and obtain the same equilibrium distribution of the Chakraborti-Chakrabarti model with arbitrary saving propensity $\lambda$,
corresponding to the dots at $\lambda = 0$ in Fig.~\ref{Fig_R1_a-VS-L},
through the use of an acceptance criterion with a suitable value of $\eta$.
This justifies at the microscopic level of the decision processes, carried out by the units before an exchange,
the use of a saving propensity $\lambda > 0$ inserted ``by hand'' in the Chakraborti-Chakrabarti model
as a convenient numerical or effective statistical procedure.

%

\section{Heterogeneous immediate-exchange model}
\label{heterogeneous}

In this section we investigate the generalization of the homogeneous
model studied above to the heterogeneous case, when each unit behaves according to a different trading criterion.
As in the homogeneous case, the presence of a heterogeneity in the parameter values can be related
to specific features of the units, such as the efficiency of the trading strategy employed by the units 
or their education level~\cite{Angle2006a}. 

Here, we limit ourselves to consider an absolute acceptance criterion.
In fact, while the equilibrium distribution of a homogeneous system is invariant under change
of the form of the (symmetrical) criterion employed,
the presence of a heterogeneity level in the system parameters is sufficient to break this invariance.
The same piece-wise linear probability function considered above is used,
but now each unit $j$ ($j = 1, \dots, N)$ is characterized by a different parameter $\eta_j$.
The corresponding acceptance probability function is given by
\begin{equation} \label{prob-4tr-het}
\begin{aligned}
q_j(\Delta x_{jk}) \!=\!
\begin{cases}
0 \, , 
&\!\!\!\mathrm{if}~  \Delta x_{jk} < - \eta_j \langle x \rangle , \\
1 + \Delta x_{jk}/(\eta_j \langle x \rangle) \, , 
&\!\!\!\mathrm{if}~ - \eta_j \langle x \rangle < \Delta x_{jk} < 0 , \\
1 \, ,          
&\!\!\!\mathrm{if}~ \Delta x_{jk} \ge 0 .
\end{cases}
\end{aligned}
\end{equation}
%
The equilibrium distribution does not have a universal shape anymore 
and depends on the specific acceptance parameters $\{\eta_i\}$ used or
on their distribution $\phi(\eta)$ that can be defined in the limit of a large number of units.

It to be noticed that heterogeneity does not concern the functional form of the amount of exchanged wealth $\Delta x_{jk}$
but enters the model through the parameters of the acceptance criterion used by each unit.

\begin{figure}[ht]
\resizebox{0.45\textwidth}{!}{
  \includegraphics{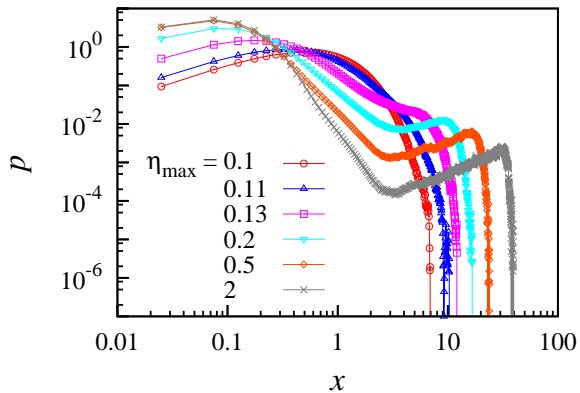}
}
\caption{\label{Fig_hetero1}
Equilibrium distribution $f(x)$ of wealth $x$ of a system of heterogeneous
units with diversified acceptance parameters $\{\eta_i\}$
uniformly distributed in the interval $(\eta_\mathrm{min}, \eta_\mathrm{max})$,
for $\eta_\mathrm{min} = 0.1$ and different values of $\eta_\mathrm{max}$.
}
\end{figure}

We have investigated various forms of distributions $\phi(\eta)$.
The equilibrium wealth distributions obtained from \emph{uniform} threshold distributions with parameters $\eta_i$ randomly extracted in an interval $(\eta_\mathrm{min},\eta_\mathrm{max})$ in general are not realistic.
An example of such equilibrium wealth distributions obtained for a fixed $\eta_\mathrm{min} = 0.1$ and different values of $\eta_\mathrm{max}$ are plotted in Fig.~\ref{Fig_hetero1}.
One can notice that when thresholds are very limited in the range (all close to $\eta = 0.1$ in the example of Fig.~\ref{Fig_hetero1})
the model is almost homogeneous and one recovers a $\Gamma$-distribution.
The presence of a significant fraction of agents $i$ with higher values of the parameter $\eta_i$ produces a depletion of the distribution in the intermediate wealth zone and an enhancement of the concentration of units both at small and large values of wealth.
In the large-wealth region one can notice the formation of a new maximum that can be interpreted as the formation of a rich class
made of the units with the smallest values of $\eta_i$'s.
To the best of our knowledge this type of maximum is not observed in real data of wealth distributions.

\begin{figure}
\resizebox{0.45\textwidth}{!}{
  \includegraphics{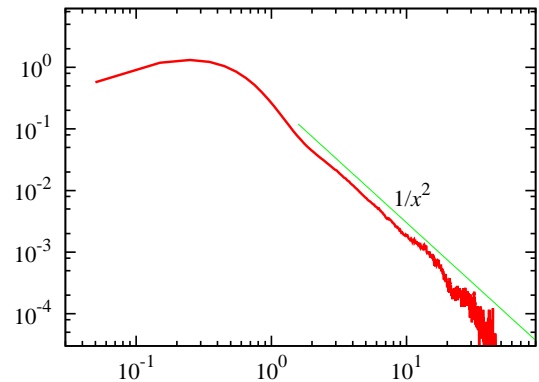}
}
\caption{\label{Fig_hetero2} 
Equilibrium distribution $f(x)$ of wealth $x$ of a system of heterogeneous
units with diversified acceptance parameters $\{\eta_i\}$: 
95\% of the units has $\eta = 2$, while the remaining 5\% has
$\{\eta_i\}$ uniformly distributed in $\eta \in (0.5, 0.7)$.
}
\end{figure}

Results reproducing a realistic shape of wealth distributions, are obtained by following a simple recipe
consisting in constructing a system in which the majority of units has a standard or low performance during trading
interactions and a small fraction of units perform much better.
This prescription was already tested in Ref.~\cite{Patriarca2006c} in the framework of the
Chakraborti-Chakrabarti model with saving propensities~\cite{Chakraborti2000a} by assigning a zero
saving propensity to 99\% of the population and higher saving propensities, uniformly distributed in $\lambda \in (0,1)$,  to the
remaining 1\% of the population.
This led to realistic shapes of wealth distribution, see Ref.~\cite{Patriarca2006c} for details.
The recipe is here adapted to the immediate-exchange model under consideration by assigning 
a (fixed) larger parameter value $\eta_i \equiv \bar{\eta} = 2$ to a majority (95\%) of the population ---
here a larger value of $\eta$ corresponds to a looser criterion used in accepting an exchange,
while the remaining 5\% of the population is assigned a set of smaller parameters $\eta_i$ uniformly extracted in the interval 
$\eta \in (0.5, 0.7)$.
The resulting equilibrium wealth distribution, shown in Fig.~\ref{Fig_hetero2}, indeed presents the stylized features of wealth
distributions, such as a mode $x_\mathrm{m} > 0$ and a Pareto power-law at larger values of wealth with a Pareto index
$p \approx 2$.

%

%

\section{Conclusions}

In the present paper we have proposed a model of wealth exchange 
aimed at overcoming some critics raised about the kinetic exchange models,
laying down firmer foundations for the models themselves,
and providing a more realistic description of a trade between two economic units.
Doing this, two main features of the original models have been modified.
First, an immediate-exchange wealth dynamics has been introduced, describing two actual and distinct wealth flows
taking place during the same interaction between two economic units, just as in an actual trade.
Models, where the total wealth is randomly reshuffled between the two units 
or which follow a uni-directional dynamics, better represent exchanges of valuables without the demand of an immediate reward, 
as, e.g., in gift economies. 
Immediate bidirectional exchanges produce more realistic shapes of wealth distribution $f(x)$, 
such that $f(0) = 0$ and having a mode larger than zero.
Furthermore, the model is made more realistic at a microeconomic level,
allowing units to decide whether to trade or not, according to some criterion based on the profit gained from the trade.
Criteria formulated symmetrically with respect to the two trading units --- but arbitrary for the rest --- 
preserve some symmetries of the model.
In the immediate-exchange models as well as in the Dr$\breve{\mathrm{a}}$gulescu-Yakovenko model, a symmetrical criterion
\emph{does not} change the equilibrium wealth distribution at all.

In the case of the Chakraborti-Chakrabarti model, when a symmetrical acceptance condition on the \emph{relative} wealth is used, 
one still finds that the equilibrium distribution is a $\Gamma$-distribution,
but with a larger value of the shape parameter $\alpha$, corresponding to a higher saving propensity $\lambda$.
This implies interesting links between different kinetic exchange models and provides at the same time a microscopic justification.
For example, the equilibrium distribution of the Chakraborti-Chakrabarti model can also be obtained from the
reshuffling dynamics of the Dr$\breve{\mathrm{a}}$gulescu-Yakovenko model  with the addition of an acceptance criterion
with suitable parameters.

The situation changes either when the criterion depends asymmetrically on the wealths of the two units (also in homogeneous
systems) or in the case of heterogeneous systems, where there is anyway a situation of asymmetry due to the fact that
each unit uses a different criterion.
In these cases the equilibrium wealth distribution depends on the specific set of criteria used by units.
The heterogeneous version of the bidirectional model presents a wide
spectrum of possible shapes for the equilibrium wealth distribution.
We have found that realistic shapes, presenting a Pareto power-law, are obtained for moderate levels of heterogeneity.

In order to provide quantitative explanations of the many regularities of and links between the various models discussed above
it would be important that future work advances also toward the analytical solution of kinetic exchange models.
Furthermore, while the ubiquitous presence of the $\Gamma$-distribution can be related 
to the validity of the Boltzmann theorem,  due to the conservation of wealth during trades,
it would be relevant to gain some information also on the analytical shapes of the equilibrium wealth distributions 
when  the Boltzmann theorem does not apply,
e.g. in the interesting Maxwell-demon-type dynamics~\cite{Apenko2014a} corresponding to cases of asymmetrical acceptance criteria
describing e.g. the ``rich gets richer'' effect.

\section*{Acknowledgement}

This work has been supported by the Estonian Science Foundation through grant no. 9462.
We are grateful to Anna Grazia Quaranta and Marcella Scrimitore for useful discussions.

\appendix   


\end{document}